\documentclass[aps,prd,twocolumn,eqsecnum,showpacs,amsmath,amssymb,superscriptaddress]{revtex4-1}

\usepackage{graphicx}
\usepackage{dcolumn}
\usepackage{bm}

\newcommand{\be}{\begin{equation}}
\newcommand{\ee}{\end{equation}}
\newcommand{\bea}{\begin{eqnarray}}
\newcommand{\eea}{\end{eqnarray}}
\newcommand{\non}{\nonumber}
\newcommand{\Non}{\nonumber\\}
\newcommand{\bseq}{\begin{subequations}}
\newcommand{\eseq}{\end{subequations}}
\newcommand{\R}{\mathcal{R}}
\allowdisplaybreaks[1]
\begin{document}


\title{General constraints on Horndeski wormhole throats}

\author{Roman Korolev}
\email{korolyovrv@gmail.com}
\affiliation{Institute of Physics, Kazan Federal University, Kremliovskaya str. 16a, Kazan 420008, Russia}

\author{Francisco S. N. Lobo}
\email{fslobo@fc.ul.pt}
\affiliation{Instituto de Astrofísica e Ci\^encias do Espa\c{c}o, Facultade de Ci\^encias da Universidade de Lisboa, Edif\'icio C8, Campo Grande, P-1749-016, Lisbon, Portugal}

\author{Sergey V. Sushkov}
\email{sergey$_\,$sushkov@mail.ru}
\affiliation{Institute of Physics, Kazan Federal University, Kremliovskaya str. 16a, Kazan 420008, Russia}

\date{\today}

\begin{abstract}
In this work, we consider the full Horndeski Lagrangian applied to wormhole geometries and present the full gravitational field equations. We analyse the general constraints imposed by the flaring-out conditions at the wormhole throat and consider a plethora of specific subclasses of the Horndeski Lagrangian, namely, quintessence/phantom fields, $k$-essence, scalar-tensor theories, covariant galileons, nonminimal kinetic coupling, kinetic gravity braiding, and the scalar-tensor representation of Gauss-Bonnet couplings, amongst others. The generic constraints analysed in this work serve as a consistency check of the main solutions obtained in the literature and draws out new avenues of research in considering applications of specific subclasses of the Horndeski theory to wormhole physics.


\end{abstract}


\maketitle

\section{Introduction}

Traversable wormholes are hypothetical shortcuts in spacetime, where the key ingredient involves the flaring-out condition at the throat \cite{Morris:1988cz,Morris:1988tu,Visser:1995cc,Lobo:2017oab}. In general relativity, through the Einstein field equation this restriction entails the violation of the null energy condition (NEC), which is defined as $T_{\mu\nu}k^{\mu}k^{\nu} \geq 0$, for {\it any} null vector $k^\mu$ \cite{Capozziello:2013vna,Capozziello:2014bqa}. However, in modified theories of gravity, it has been shown that the NEC can be satisfied for normal matter threading the wormhole throat, and it is the higher order curvature terms that sustain the wormhole geometry \cite{Harko:2013yb}.
In fact, wormhole physics has been extensively explored in modified theories of gravity, such as in $f(R)$ gravity \cite{Lobo:2009ip}, extended theories with a nonminimal curvature-matter coupling \cite{Garcia:2010xb,MontelongoGarcia:2010xd}, scalar-tensor theories with nonminimal derivative coupling \cite{Sushkov:2011jh,Korolev:2014hwa}, hybrid metric-Palatini gravity \cite{Capozziello:2012hr} and its generalized version \cite{Rosa:2018jwp}, and higher-dimensional theories \cite{Zangeneh:2015jda,Mehdizadeh:2015dta,Mehdizadeh:2015jra,Zangeneh:2014noa,Mehdizadeh:2016nna}, amongst many other theories (we refer the reader to Ref. \cite{Lobo:2017oab} for a recent review). These extended theories of gravity admit an equivalent scalar-tensor representation.

Indeed, scalar fields are popular building blocks used to construct physical theories and are appealing as such fields are ubiquitous in theories of high energy physics beyond the standard model. Given the large number of models, the question arises how we should study and compare them in a unified manner. A particularly useful tool in this direction is the realisation that all these classes of models are special cases of the most general Lagrangian which leads to second order field equations, namely, the Horndeski Lagrangian \cite{Horndeski:1974wa}, which was recently rediscovered \cite{Deffayet:2011gz}.
The Horndeski action can be given by 
\be
{\cal S}=\int{}\;d^4x\sqrt{-g}\sum^5_{i=2}{\cal L}_i
+{\cal S}_M[g_{\mu\nu},\chi_M]\,,
\label{action}
\ee
where $g$ is the determinant of metric tensor $g_{\mu\nu}$, 
${\cal S}_M$ is the matter action, in which $\chi_M$ collectively denotes all matter fields.
The Lagrangians ${\cal L}_{i}$ are defined by
\bea
{\cal L}_2 &=& G_2(\phi,X)\,,
	\non\\
{\cal L}_3 &=& - G_3(\phi,X)\Box\phi\,,
	\non\\
{\cal L}_4 &=& G_{4}(\phi,X) {R} +G_{4,X}(\phi,X)
\left[ (\square \phi )^{2}-(\nabla_\mu \nabla_\nu \phi)^2  \right] 
	\non\\
{\cal L}_5 &=& G_{5}(\phi,X) G_{\mu\nu}\,\nabla^\mu \nabla^\nu \phi -\frac{1}{6}G_{5,X}\times
\Non
&& \times \left[\left( \Box \phi \right)^3 -3 \Box \phi (\nabla_\mu \nabla_\nu \phi)^2 + 2\left(\nabla_\mu \nabla_\nu \phi \right)^3 \right],
\eea
respectively, where ${R}$ is the Ricci scalar and $G_{\mu\nu}$ is the Einstein tensor; the factors $G_{i}$ ($i=2,3,4,5$) are arbitrary functions of the scalar field $\phi$ and the canonical kinetic term,
$X=-\frac{1}{2}\nabla^\mu\phi \nabla_\mu\phi$. We consider the definitions $G_{i,X}\equiv \partial G_i/\partial X$, $(\nabla_\mu \nabla_\nu \phi)^2=\nabla_\mu \nabla_\nu \phi  \,\nabla^\nu \nabla^\mu \phi$, and $\left(\nabla_\mu \nabla_\nu \phi \right)^3= \nabla_\mu \nabla_\nu \phi  \,\nabla^\nu \nabla^\rho \phi \, \nabla_\rho \phi  \nabla^\mu \phi$. Furthermore, we assume that the matter fields $\chi_M$ are minimally coupled to gravity.

Note that by choosing the functions $G_i$ appropriately, one may reproduce any second-order scalar-tensor tensor theory. For instance, the $G_2$ term is used in $k-$essence \cite{Chiba:1999ka,ArmendarizPicon:2000dh}, and the $G_3$ term was explored in the context of kinetic gravity braiding \cite{Deffayet:2010qz}. One may also consider theories in which the scalar field is nonminimally coupled to the Ricci scalar ${R}$ in the form  $G_4(\phi) {R}$ \cite{Fujiibook}, where a representative example is Brans-Dicke (BD) theory \cite{Brans} with a scalar potential $V(\phi)$. The specific assumption of $G_4={\rm constant}$ provides the Hilbert-Einstein term. The choices of $G_4=X$ or $G_5=-\phi$ have been used in nonminimal couplings of the form $G_{\mu\nu}\,\nabla^\mu \nabla^\nu \phi$ \cite{Saridakis:2010mf,Germani:2010gm}. We refer the reader to \cite{Kase:2018aps} for a plethora of specific cases. 
These theories all belong to a subclass of more general second-order scalar-tensor theories denoted by Horndeski theories \cite{Horndeski:1974wa}.

An interesting application of scalar-tensor theories is in wormhole physics, where a wide variety of solutions have been obtained in the literature \cite{Barcelo:2000zf,Barcelo:1999hq,Bronnikov:2010tt,Bronnikov:2002sf,Matos:2005uh,Shaikh:2016dpl,Kashargin:2008pk}, especially related to the stability issues \cite{Gonzalez:2008wd,Gonzalez:2008xk,Kanti:2011jz,Kanti:2011yv,Dzhunushaliev:2013lna}. These (and other) solutions are specific sub-classes of the Horndeski action  (\ref{action}), so it is important to consider the most general conditions that are needed to obtain wormhole geometries in Horndeski theories. Thus, in this work, we consider the analysis restricted to the wormhole throat and analyse a wide variety of subclasses of the Horndeski Lagrangian. This proves to be extremely useful as it serves as consistency checks for the solutions obtained in the literature and paves the way for new avenues of research related to subclasses of Horndeski wormhole solutions.

This work is organised in the following manner: In Sec. \ref{S:wormholethroat}, we present the wormhole metric and the general constraints of the full Horndeski Lagrangian at the wormhole throat. 
In Sec. \ref{S:specific_cases}, we consider specific subclasses of the Horndeski theory, namely, quintessence/phantom fields, $k$-essence, scalar-tensor theories, covariant galileons, nonminimal kinetic coupling, kinetic gravity braiding, and the scalar-tensor representation of Gauss-Bonnet couplings, amongst others. In Sec. \ref{S:conclusion}, we conclude and discuss our results.

In addition to this, as the gravitational field equations for the full Horndeski theory are extremely lengthy, we opt to present these in  Appendix \ref{fieldeqs}. The field equations at the throat are written in Appendix \ref{A:fieldeqs_throat}, which are then used to deduce the most general flaring-out condition for Horndeski wormholes, in terms of the scalar field $\phi$, the kinetic term $X$, the factors  $G_{i}$ and their derivatives, and presented in Appendix \ref{flareout_throat}.

\section{General analysis at the wormhole throat}\label{S:wormholethroat}

\subsection{Metric and flaring-out condition}

Consider a static and spherical symmetric configuration in the theory (\ref{action}). In this case the spacetime metric can be taken as follows:
\begin{equation} \label{metric}
ds^2=-A(u)dt^2+A^{-1}(u)du^2+r^2(u)d\Omega^2,
\end{equation}
where $d\Omega^2=d\theta^2+\sin^2\theta d\varphi^2$ is the linear element of the unit sphere, and the metric functions $A(u)$, $r(u)$ and the scalar field $\phi$ are functions of the radial coordinate $u$.
Here the radial coordinate lies in the range $u \in (-\infty, + \infty)$, so that two asymptotically flat regions exist, i.e., $u \rightarrow \pm \infty$, and are connected by the throat. In addition to this, the function $r(u)$ possesses a global positive minimum at the wormhole throat $u=u_0$, which one can set at $u_0=0$, without a loss of generality. 

The wormhole throat is defined as $r_0={\rm min}\{r(u)\}=r(0)$. In order to avoid event horizons and singularities throughout the spacetime, one imposes that the function $A(u)$ is positive and regular everywhere. 
Taking into account these restrictions, namely, the necessary conditions for the minimum of the function imposes the flaring-out conditions, translated as 
\be
r_0'=0\,, \qquad  r_0'' > 0\,.  
\label{flareout}
\ee

As the metric function $A(u)$ is positive and regular for $\forall u$, it is useful to analyse its first and second derivatives at the throat $u=0$. Thus, $A_0$ is a free parameter, as is $A_0'$, which for simplicity, we impose hereafter $A_0'=0$. Now, the sign of $A_0''$ determines the type of extrema of $A(u)$, i.e., it is a minimum if $A_0''>0$ and a maximum if $A_0''<0$.
This implies that the maximum (minimum) of $A(u)$ corresponds to a maximum (minimum) of the gravitational potential, so that in the vicinity of a maximum (minimum) the gravitational force is repulsive (attractive). Thus, the wormhole throat possesses a repulsive or an attractive nature that depends on the sign of $A_0''$.


\subsection{Generic constraints at the wormhole throat}

We present the full field equations for the Horndeski action (\ref{action}), using the metric (\ref{metric}), in Appendix \ref{fieldeqs}. Thus, taking into account the field equations (\ref{tt})--(\ref{theta}), evaluated at the throat, and using the condition $A_0'=0$ (see discussion above), one obtains restrictions for the wormhole geometry.
More specifically, setting $r’_0=0$ at the throat, we obtain a set of linear algebraic equations for the second 
derivatives $r''_{0}$, $A''_{0}$, $\phi''_{0}$. In this work, we use the three components of the gravitational 
field equations (\ref{tt})--(\ref{theta}) and the scalar field equation (\ref{App:scalareq}) presented in  
Appendix \ref{fieldeqs}. 

The $(rr)$-component (\ref{rr}) of the field equations is of first order and represents a constraint on initial conditions, and is given by 
\be
\frac{1}{r_0^2}= - \frac{1}{2}\; \frac{G_2  +  A_0 \phi_0'^2 \left(G_{2,X} -G_{3,\phi} \right)}
{G_4 + A_0 \phi_0'^2 \left( G_{4,X} - \frac{1}{2} G_{5,\phi}\right) } \Bigg|_{u_0} \,. \label{ttthroat}
\ee
This condition places an additional constraint on the wormhole geometry, as $1/r_0^2 > 0$.
Note that, taking into account the metric (\ref{metric}), the kinetic term at the throat is negative and takes the form $X_0=-\frac{1}{2}A_0 \phi_0'^2 <0$. 

Furthermore, in order to obtain the flaring-out condition, we have to resolve the linear algebraic equations with respect to $r''_0$, which places further constraints on the wormhole geometry through the flaring-out condition (\ref{flareout}).
For this purpose, we write out the field equations at the throat, presented in Appendix \ref{fieldeqs}, by taking into account the conditions $r'_0=0$ and $A'_0=0$, so that Eqs. (\ref{tt}), (\ref{theta}) and the scalar field equation (\ref{App:scalareq}) reduce to Eqs. (\ref{tt_thr})--(\ref{scalar_throat}) in Appendix \ref{A:fieldeqs_throat}.
Finally, eliminating the terms $A''_0$ and $\phi''_0$, one finally arrives at the most general flaring-out condition for Horndeski wormholes, solely in terms of the scalar field $\phi$, the kinetic term $X$, the factors  $G_{i}$ and their derivatives, given by Eq. (\ref{rrthroat}), which due to its extremely lengthy and messy form is presented in Appendix \ref{flareout_throat}. We then refer to this condition, when analysing specific subclasses of the Horndeski action.

Note that these relations are constrained by the imposition $r_0^2>0$ and the flaring-out condition, $r''_0>0$, at the throat. Thus, in order to be a wormhole solution, these equations impose tight restrictions on the spacetime geometry. The strategy to follow is take into account these conditions in order to analyse and serve as a consistency check on specific solutions obtained in the literature. In addition to this, one can obtain generic impositions to obtain novel solutions and which may trace out new avenues of research in wormhole physics, in the context of subclasses of Horndeski theories.

\section{Subclasses of Horndeski theory}\label{S:specific_cases}

\subsection{Quintessence/phantom fields}

Recent observations suggest that a large fraction of the energy density of the universe has negative pressure, where a possible explanation is in the form of a scalar field slowly evolving down a potential, denoted quintessence \cite{quin4,quin6, quin7}. The latter possesses a positive kinetic energy, however, phantom scalar fields with a negative kinetic energy \cite{Bronnikov:2001tv} have also been considered.

For the specific case of quintessence/phantom fields, consider the following functions
\be 
 G_2= \epsilon X-V(\phi)\,, \quad G_4 = \frac{1}{16\pi} \,, \quad G_3=G_5=0 \,,
\ee 
where $\epsilon=\pm 1$. Note that the case $\epsilon=+1$ corresponds to the standard canonical term \cite{quin4,quin6, quin7}, and $\epsilon=-1$ to a phantom field which possesses a negative kinetic energy \cite{Bronnikov:2001tv}. In fact the phantom field rolls up the potential due to the negative kinetic energy, so that if the potential is unbounded from above, the energy density tends to infinity. 

Recall that taking into account the metric (\ref{metric}), the kinetic term at the throat is negative, i.e., $X_0=-\frac{1}{2}A_0 \phi_0'^2 <0$. Thus, Eqs. (\ref{ttthroat}) and (\ref{rrthroat}) reduce to 
%
%
\bea
\frac{1}{r_0^2}&=& 8\pi \left( \epsilon X_0 + V_0 \right) \,, \label{GRtt} \\
\frac{r_0''}{r_0} &=& \frac{8 \pi \epsilon  X_0}{A_0} \,,  \label{GRrr}
\eea
respectively, from which we verify that Eq. (\ref{GRrr}) is consistent with the flaring-out condition, i.e., $r_0''>0$, only if $\epsilon <0$, corresponding to a phantom field. This reproduces the well-known result that a wormhole solution in general relativity is only permitted with a minimally coupled phantom scalar field with negative kinetic energy \cite{Bronnikov:2001tv,Sushkov:2005kj,
Lobo:2005us}. Condition (\ref{GRtt}) imposes that $V_0> -A_0 \phi_0'^2/2$ (for $\epsilon = -1$). 
These results are consistent with those presented in \cite{Bronnikov:2001tv,Sushkov:2005kj,
Lobo:2005us,Lobo:2005yv,Bronnikov:2006pt,Garattini:2007ff,Cataldo:2008pm,Gonzalez:2009cy,
Bronnikov:2012ch,Lobo:2012qq,Kleihaus:2014dla}. 

In fact, it has been suggested that a possible candidate for the present accelerated expansion of the Universe is ``phantom energy'' \cite{Caldwell:2003vq}, which 
possesses an equation of state of the form $\omega = p/\rho <-1$, where $\rho$ is the energy density and $p
$ the pressure. This consequently violates the null energy condition, which is the fundamental ingredient to 
sustain traversable wormholes, so that this cosmic fluid presents us with a natural scenario for the existence 
of these exotic geometries.

\subsection{$k$-essence}

As quintessence is based on a canonical scalar field with a potential, it is known that scalar fields with non-canonical kinetic terms appear in high energy physics. This motivates the presence of an arbitrary function of the scalar field and the kinetic term in the gravitational Lagrangian. Therefore, one may consider the following functions, which generalize the case studied above 
\be 
 G_2= K(\phi,X)\,, \quad G_4 = \frac{1}{16\pi}\,, \quad G_3=G_5=0 \,,
\ee 
so that Eqs. (\ref{ttthroat}) and (\ref{rrthroat}) take the form
%
\bea
\frac{1}{r_0^2}&=& -8\pi \left( K_0  - 2 X_0 K_{X_0} \right) \,,  \label{kGRtt} \\
\frac{r_0''}{r_0} &=& \frac{8\pi X_0 K_{X_0}}{A_0} \,, \label{kGRrr}
\eea
respectively.

Thus, Eqs. (\ref{kGRtt}) and (\ref{kGRrr}) impose the following conditions at the wormhole throat
\bea \label{whconditions}
K_0 -2X_0 K_{X_0} < 0 \,, \quad K_{X_0} <0 \,.
\eea
(The first condition can be written in the following form $K_0 +A_0 \phi_0'^2 K_{X_0} < 0$).
More specifically, $K_{X_0} < 0$, imposes that $K_0 <  A_0 \phi_0'^2 |K_{X_0}|$.
These conditions are consistent with those presented in Ref. \cite{Bronnikov:2015aam}, where static and spherically symmetric configurations in the context of $k$-essence theories defined by a function depending solely on the kinetic term, i.e.,  $K=K(X)$, were presented. In fact, a no-go theorem was proved, claiming that a possible black-hole-like Killing horizon of finite radius cannot exist if the function $K(X)$ is required to have a finite derivative $dK/dX$.

As a specific example, consider the case of a ghost condensate model \cite{ArkaniHamed:2003uy}, given by the function $K=-X+X^2/M^4$, so that conditions (\ref{whconditions}) impose 
\bea 
X_0 - \frac{3}{M^4} X_0^2 < 0 \,, \quad -1+\frac{2}{M^4}X_0 < 0 \,,
\eea
which taking into account that $X_0 < 0$, are automatically satisfied.

One may also consider actions arising from low-energy effective string theory, which gives rise to higher-order derivative terms coming from loop corrections to the tree-level action \cite{Gasperini:2002bn}. For instance, consider the specific case: $K(\phi,X)=\bar{K}(\phi)X + L(\phi) X^2$, so that $K_X=\bar{K} + 2L X$, and conditions (\ref{whconditions}) impose:
%
\be
\bar{K} (\phi_0) <  -2X_0 L(\phi_0) \,, \quad \bar{K}(\phi_0) <  -3 X_0 L(\phi_0)  \,,
\ee
where the bounds essentially depend on the signs of the functions $\bar{K}(\phi_0)$ and $L(\phi_0)$.

\subsection{Scalar-tensor theories}

Scalar fields have a long history in gravitation, starting with Brans-Dicke theory \cite{Brans}, in which gravity is mediated by the scalar field and the metric tensor field. In fact, modified theories of gravity may be written in a scalar-tensor representation, by introducing specific Legendre transformations, which motivates the further analysis of specific cases of scalar-tensor theories in more detail. Consider the general nonminimally coupled theories given by the couplings \cite{Faraoni:2004pi,Gas92,Damour1,Chiba99,Bartolo99,Boi00}:
\be
G_2=\omega (\phi)X-V(\phi), \qquad  G_4=F(\phi), \qquad G_3=G_5=0\,,
\ee
so that Eqs. (\ref{ttthroat}) and (\ref{rrthroat}) take the form
%
\be
\frac{1}{r_0^2}=\frac{1}{2F_0}\left( V_0 +\omega X_0 \right)\,, 
\ee
and
\bea
\frac{r''_0}{r_0}= \Big\{\omega XF\left( \omega + 2F''  \right) +\left( \omega X + V  \right) F'^2 
	 \nonumber \\
	+FF' \left[  X ( \omega- \omega') + V - V'  \right] \Big\}\Big|_{u_0}\Big/
	\nonumber \\
	\left[  2AF \left(  \omega F + FF' +2F'^2   \right)  \right]\big|_{u_0}\,,
\eea
respectively.

In order to have wormhole geometries, as mentioned above, these quantities are imposed to be positive. We will analyse specific cases below, namely, Brans-Dicke theory with a potential, and the scalar-tensor representations of several modified theories of gravity.

\subsubsection{Brans-Dicke theory}

Wormholes physics has been extensively explored in the context of Brans-Dicke theory \cite{Agnese:1995kd,Anchordoqui:1996jh,Nandi:1997en,Nandi:1997mx,Bhattacharya:2009rt,
Lobo:2010sb,MontelongoGarcia:2011ag,Sushkov:2011zh,Papantonopoulos:2019ugr}. Here, we consider the most general conditions for the existence of these exotic geometries.
In Brans-Dicke theory \cite{Brans} with the scalar potential 
$V(\phi)$, we have
\bea
G_2=\frac{1}{16\pi}\left(\frac{\omega_{\rm BD}}{\phi}X-V(\phi)\right)\,,
	\non\\  
	G_4=\frac{1}{16\pi}\phi\,,\quad G_3=G_5=0\,.
\label{BDaction}
\eea
In the limit that $\omega_{\rm BD} \to \infty$, we recover GR with a quintessence scalar field. 

For this case, Eqs. (\ref{ttthroat}) and (\ref{rrthroat}) take the form
\bea
\frac{1}{r_0^2}&=&\frac{1}{2\phi_0^2}\left(\omega_{\rm BD} X_0 +\phi_0 V_0 \right)\,, \label{BDGRtt} \\
\frac{r''_0}{r_0}&=& \frac{\omega_{\rm BD} X_0}{2A_0 \phi_0^2} + \frac{V_0(1 + \phi_0)-V_0' \phi_0}{2A_0 \phi_0 (2+\omega_{\rm BD} +\phi_0)} \,, \label{BDGRrr}
\eea
respectively. Note that in the absence of the potential, $V=0$, conditions (\ref{BDGRtt}) and (\ref{BDGRrr}) impose that $\omega_{\rm BD}<0$, which is consistent with the literature. However, considering a non-zero potential alleviates this restriction, where inequality (\ref{BDGRtt}) imposes a general condition on the value of the potential at the throat given by $\phi_0 V_0 > -\omega_{\rm BD} X_0 $ (note that this relaxes the restriction $\omega_{\rm BD}<0$). On the other hand, inequality (\ref{BDGRrr}) imposes an inequality on the derivative of the potential, assuming that $2+\omega_{\rm BD} +\phi >0$, given by
\begin{equation}
V_0' < \frac{(2+\omega_{\rm BD} +\phi)\omega_{\rm BD}X_0}{\phi_0^2} + \frac{V_0(1+\phi_0)}{\phi_0}\,.
\label{BDdpot}
\end{equation}

Below, we consider specific cases of modified theories of gravity, that can be represented as particular cases of Brans-Dicke theory.

\subsubsection{$f(R)$ gravity: metric formalism}

An extension of general relativity that has recently been explored in detail is $f(R)$ gravity, in order to explain the late-time cosmic acceleration \cite{Sotiriou:2008rp,Lobo:2008sg}.
The action of $f(R)$ gravity is given by  
\be
{\cal S}_{\rm H}=\frac{1}{16\pi} \int d^4x \sqrt{-g}\,f(R)\,,
\label{fRaction}
\ee
where $f(R)$ is an arbitrary function of $R$. The metric $f(R)$ gravity, which corresponds to the variation 
of (\ref{fRaction}) with respect to $g_{\mu \nu}$, is represented by the following choices of the Lagrangian
\be
G_2=-\frac{1}{16\pi} (R F-f),\quad G_4=\frac{1}{16\pi} \,F\,,\quad G_3=G_5=0\,,
\label{fRcase}
\ee
where $F(R) \equiv \partial f/\partial R$. 
This corresponds to the Jordan frame representation of the action of a Brans–Dicke theory with $\omega_{\rm BD}= 0$ and we define the scalar potential as $V= (RF-f)$ \cite{Ohanlon,Chiba03}. Here the scalar degree of freedom $\phi=F(R)$ arises from the gravitational sector.

Note, however, that it has been argued in the literature that traversable wormhole geometries are only valid in the interval $-3/2< \omega_{\rm BD} < -4/3$ \cite{Nandi:1997en}, consequently apparently excluding $f(R)$ gravity wormholes, which are equivalent their Brans-Dicke counterparts, with $\omega_{\rm BD}=0$. However, it was shown that this referred interval is only valid for a specific choice of an integration constant of the field equations derived on the basis of a post-Newtonian weak field approximation, and there is no reason for it to hold in the presence of compact objects with strong gravitational fields \cite{Lobo:2010sb}.

For this case, Eqs. (\ref{ttthroat}) and (\ref{rrthroat}) take the form
\bea
\frac{1}{r_0^2}&=&\frac{V_0}{2\phi_0}\,, \label{fRmettt} \\
\frac{r''_0}{r_0}&=& \frac{V_0(1+\phi_0 )-V_0' \phi_0}{2A_0 \phi_0 (2 +\phi_0)} \,, \label{fRmetrr}
\eea
respectively.  Here, inequality (\ref{fRmettt}) imposes the following generic restriction $V_0/\phi_0 >0$, which may be interpreted as a constraint on the potential at the throat. Inequality (\ref{fRmetrr}) imposes a generic condition on the derivative of the potential at the throat, assuming that $2 +\phi_0 > 0$, given by
\begin{equation}
V_0' <  \frac{V_0(1+\phi_0)}{\phi_0}\,.
\end{equation}
Note that this restriction is consistent with considering $\omega_{\rm BD}=0$ in inequality (\ref{BDdpot}).

\subsubsection{$f(R)$ gravity: Palatini approach}

The Palatini $f(R)$ gravity, which corresponds to the variation of (\ref{fRaction}) with respect to $g_{\mu \nu}$ and the connection, in the scalar-tensor representation corresponds to a Brans-Dicke theory with the BD parameter $\omega = -3/2$ \cite{Olmo:2011uz}. Thus, Eqs. (\ref{BDGRtt}) and (\ref{BDGRrr}) reduce to 
%
\bea
\frac{1}{r_0^2}&=&\frac{ 2\phi_0 V_0-3 X_0 }{4\phi_0^2}\,, \label{PalatiniGRtt} \\
\frac{r''_0}{r_0}&=& - \frac{3 X_0}{4A_0 \phi_0^2} + \frac{V_0(1 + \phi_0)-V_0' \phi_0}{A_0\phi_0 (1+2\phi_0 )} \,, \label{PalatiniGRrr}
\eea

Condition (\ref{PalatiniGRtt}) provides $\phi_0 V_0 > 3 X_0/2$, while inequality (\ref{PalatiniGRrr}) imposes the following generic constraint on the derivative of the potential evaluated at the throat, assuming that $1+2\phi_0 >0$
\begin{equation}
V_0' < -\frac{3(1+2\phi_0)X_0}{4\phi_0^2} + \frac{V_0(1+\phi_0)}{\phi_0}\,.
\label{Palatinicond}
\end{equation}

In fact, in this context, nontrivial wormhole topologies in Planck-suppressed quadratic extensions of General Relativity (GR) formulated in the Palatini formalism have been explored and the physical significance of curvature divergences in theory and the topology change issue have been analysed in the literature \cite{Lobo:2013vga,Lobo:2014fma,Lobo:2014zla,Bejarano:2016gyv}. This study supports the view that spacetime could have a foam-like microstructure pervaded by wormholes generated by quantum gravitational effects.

\subsubsection{Hybrid metric-Palatini theory}

It has been established that both metric and Palatini versions of $f(R)$ theories of gravity have interesting features but also manifest severe and different downsides. A hybrid combination of theories, containing elements from both these two formalisms, turns out to be also very successful accounting for the observed phenomenology and is able to avoid some drawbacks of the original approaches \cite{Harko:2011nh,Capozziello:2012ny,Capozziello:2012hr,Capozziello:2015lza}. 
More specifically, this approach consists of adding to the Einstein-Hilbert Lagrangian an $f(R)$ term constructed {\it a la} Palatini \cite{Harko:2011nh}. Using the respective dynamically equivalent scalar-tensor representation, it has been shown that the theory passes the Solar System observational constraints even if the scalar field is very light. This implies the existence of a long-range scalar field, which is able to modify the cosmological and galactic dynamics, but leaves the Solar System unaffected.

The action of the hybrid metric-Palatini theory is given by \cite{Harko:2011nh}:
\begin{equation} \label{eq:S_hybrid}
S= \frac{1}{16\pi}\int d^4 x \sqrt{-g} \left[ R + f({\R})\right] +S_m \ ,
\end{equation}
where $S_m$ is the matter action, $R$ is the Einstein-Hilbert term, ${\R} \equiv  g^{\mu\nu}{\R}_{\mu\nu} $ is the Palatini curvature,  and ${\R}_{\mu\nu}$ is defined in terms of an independent connection given by
${\R}_{\mu\nu} \equiv \hat{\Gamma}^\alpha_{\mu\nu , \alpha} - \hat{\Gamma}^\alpha_{\mu\alpha , 
\nu} + \hat{\Gamma}^\alpha_{\alpha\lambda}\hat{\Gamma}^\lambda_{\mu\nu} 
-\hat{\Gamma}^\alpha_{\mu\lambda}\hat{\Gamma}^\lambda_{\alpha\nu}$.

The action (\ref{eq:S_hybrid}) may be expressed as the following scalar-tensor theory
\begin{equation} \label{eq:S_scalar2}
S=\int \frac{d^4 x \sqrt{-g} }{16\pi}\left[ (1+\phi)R +\frac{3}{2\phi}\partial_\mu \phi
\partial^\mu \phi -V(\phi)\right]+S_m ,
\end{equation}
which differs from $w=-3/2$ Brans-Dicke theory in the coupling of the scalar to the
curvature, which in the $w=-3/2$ theory is $\phi R$. 

Thus, the Horndeski $G_i$ factors are given by 
\bea
G_2=\frac{1}{16\pi}\left(-\frac{3}{2\phi}X-V(\phi)\right)\,,
	\non\\  
	G_4=\frac{1}{16\pi}(1+\phi)\,,\quad G_3=G_5=0\,,
\label{hybridBDaction}
\eea
so that Eqs. (\ref{ttthroat}) and (\ref{rrthroat}) take the form
\bea
\frac{1}{r_0^2}&=&\frac{ 2\phi_0 V_0-3 X_0 }{4\phi_0(1+\phi_0)}\,, \label{hybridGRtt} 
	\\
\frac{r''_0}{r_0}&=& \Big\{4\phi_0^2 V_0(2+\phi_0) -4V'_0 \phi_0^2(1+\phi_0) 
	\nonumber \\
&&\quad +3X_0 \left[1-\phi_0(3+2\phi_0)\right] \Big\} \big/
	\nonumber  \\
&& \qquad [4A_0\phi_0(2\phi_0^3 + 5\phi_0^2 -3)]
\,. \label{hybridGRrr}
\eea

Assuming that $\phi_0(1+\phi_0)>0$, then condition (\ref{hybridGRtt}) imposes $\phi_0 V_0-3 X_0/2 >0$ in order to have wormhole solutions \cite{Bronnikov:2019ugl}, and the positivity of condition (\ref{hybridGRrr}) may  be interpreted as a condition of the derivative of the potential.

\subsection{Nonminimal kinetic coupling}

One of the simplest Lagrangians in the Horndeski theory contains a nonminimal kinetic coupling of a scalar field to the curvature. In fact, cosmological applications have been explored in the literature  \cite{Sushkov:2009hk,Sushkov:2012za,Granda:2010ex,Granda:2011zk}.
Relative to wormhole physics, solutions with a nonminimal kinetic coupling were studied in \cite{Sushkov:2011jh,Korolev:2014hwa}. More specifically, general solutions describing asymptotically flat traversable wormholes were obtained by means of numerical methods \cite{Korolev:2014hwa}, and particular exact wormhole solutions in an analytical form have been found by using the Rinaldi method \cite{Sushkov:2011jh}. 

Consider the functions
\be
 G_2= \epsilon X-V(\phi)\,, \quad G_3=0\,, \quad  G_4 = \frac{1}{16\pi}  \quad G_5=\frac{1}{2} \eta \phi \,,
\ee
so that Eqs. (\ref{ttthroat}) and (\ref{rrthroat}) take the form
%
\bea
 \frac{1}{r_0^2}&=&  \frac{8\pi  \left(\epsilon X_0+V_0 \right)}{1+8\pi \eta X_0}  \,, \\
 \frac{r_0''}{r_0} &=& \frac{8\pi\,X_0 \left(\epsilon -  8\pi \eta V_0 \right)}{A_0\left(1+8\pi \eta X_0\right)^2}\,,
\eea
which is consistent with the results extensively explored in Ref. \cite{Sushkov:2011jh}.
Indeed, rather than analyse these results, we refer the reader to \cite{Sushkov:2011jh,Korolev:2014hwa} for more details.

\subsection{Kinetic gravity braiding (KGB)}
 
A large class of scalar-tensor models with interactions containing the second derivatives of the scalar field, but not leading to additional degrees of freedom, has been introduced. These models exhibit peculiar features, such as an essential mixing of scalar and tensor kinetic terms \cite{Deffayet:2010qz}, and have been denoted by kinetic braiding. It is interesting that this braiding essentially causes the scalar stress tensor to deviate from the perfect-fluid form \cite{Pujolas:2011he}, and in particular, in cosmology these models possesses a rich phenomenology. In fact, the late-time asymptotic is a de Sitter state, and the scalar field can exhibit a phantom behaviour that is able to cross the phantom divide with neither ghosts nor gradient instabilities.  

For the kinetic gravity braiding, consider the functions
\be
 G_2= \epsilon X-V(\phi)\,, \quad G_3 \neq 0\,,  \quad
 G_4 = \frac{1}{16\pi} \,, \quad G_5=0\,,
\ee
 so that Eqs. (\ref{ttthroat}) and (\ref{rrthroat}) take the form
\be
 \frac{1}{r_0^2}= 8\pi  \left[X_0  \left( \epsilon - 2G_{3,\phi} \right) + V_0  \right]  \,, 
 	\label{ttKGB} 
\ee
\bea
 \frac{r_0''}{r_0} &=& \frac{8\pi  X_0}{A_0}  \Big[ 2XG_{3,\phi X}(2G_{3,\phi}-\epsilon)  - G_{3,X} G_{3,\phi\phi}(\phi+4 X)
	\nonumber \\	 
	&&-16\pi X G_{3,X}^2 (2XG_{3,\phi}-\epsilon X-V) + (2 G_{3,\phi}-\epsilon)^2
	\nonumber \\
	&&+G_{3,X} (2XG_{3,\phi}-\epsilon X-V +V')
  \Big]\Big|_{u_0} \Big/ 
  \Big[\epsilon - 2 G_{3,\phi} 
	\nonumber  \\  
	&& - 2 G_{3,\phi X} X  + X G_{3,X} (48\pi X G_{3,X}-1) \Big]\Big|_{u_0}
  \,, \label{rrKGB}
\eea
respectively.

From Eq. (\ref{ttKGB}), one finds that the general condition $X_0  \left( \epsilon - 2G_{3,\phi} \right) + V_0 < 0$ is imposed. However, one cannot extract much information from inequality (\ref{rrKGB}), and we will resort to specific cases. For instance, consider the case of $G_3=\lambda g(\phi)$, so that Eqs. (\ref{ttKGB}) and (\ref{rrKGB}) reduce to 
\bea
 \frac{1}{r_0^2} &=& - 8\pi X_0  \left( 2g' \lambda - \epsilon - V_0  \right)  \,, 
 	\label{ttKGB2}  \\
 \frac{r_0''}{r_0} &=& - 8\pi X_0  \left( 2g' \lambda - \epsilon   \right)  \,, 
    \label{rrKGB2}
\eea
which impose the conditions $(2g' \lambda - \epsilon - V_0) > 0$ and $(2g' \lambda - \epsilon) > 0$. For the specific simple linear case of $g(\phi)=\phi$ and with zero potential $V=0$, one has the generic condition imposed on the wormhole throat $2\lambda > \epsilon$

For the specific case of functions solely on the kinetic term, namely, $G_2=K(X)$ and $G_3=G_3(X)$ \cite{Kobayashi:2010cm}, for instance, taking into account $G_2=-X+\lambda X^2$ and $G_3= \eta X$, the conditions (\ref{ttKGB}) and (\ref{rrKGB}) reduce to the following:
\be
 \frac{1}{r_0^2} = - 8\pi X_0  \left( 1-3\lambda X_0  \right)  , 
 	\label{ttKGB3}  
 	\ee
and
 	\bea
&& \frac{r_0''}{r_0} = - \frac{8\pi X_0}{A_0}  \Big[ 1 + \eta X_0 \left(1-16\pi \eta  X_0 \right) \left( 1-3\lambda X_0  \right)    - 4\lambda X_0  
	\nonumber \\
	&& \quad \times  \left( 2-3\lambda X_0  \right)  \Big] \Big/
	\left( 1 + \eta X_0 -6\lambda X_0  - 48\pi \eta^2 X_0^2 \right), 
    \label{rrKGB3}
\eea
respectively.
Taking into account $X_0<0$, condition (\ref{ttKGB3}) imposes the generic constraint $3\lambda X<1$. Note that for $\lambda \geq 0$, condition (\ref{ttKGB3}) is automatically satisfied, and for the specific case of  $\lambda = 0$, condition (\ref{rrKGB3}) simplifies to:
  \bea
 \frac{r_0''}{r_0} &=& - \frac{8\pi X_0}{A_0} \left( \frac{  1 + \eta X_0  - 16\pi \eta^2 X_0^2}{ 1 + \eta X_0  - 48\pi \eta^2 X_0^2 }\right)
 \,,
    \label{rrKGB3b}
\eea
which imposes constraints of the factor $\eta$, so that the term in parenthesis should be positive; note that $\eta=0$ is automatically satisfied, reduces to the phantom case with zero potential.
 
Wormhole geometries in the context of the kinetic gravity braiding has been largely unexplored in the literature, and the authors of the present work are currently analysing several lines of research in this direction.

\subsection{Covariant galileons}

In the context of the original Galileons, the field equations are invariant under the  shift $\partial_{\mu} \phi \to \partial_{\mu} \phi+b_{\mu}$ in Minkowski spacetime \cite{Nicolis}. However, in curved spacetime, the construction of covariant Galileon Lagrangians \cite{Galileons} maintains the equations of motion up to second order, and recovers the Galilean shift symmetry in the Minkowski limit.
Indeed, covariant Galileons are characterized by the functions
\bea
&&G_2=\beta_1 X-m^3 \phi\,,\qquad G_3=\beta_3 X\,,
	\non \\
&&G_4=\frac{M_{\rm pl}^2}{2}+\beta_4 X^2\,, \qquad
G_5=\beta_5 X^2\,,
\eea
where $m$ and $\beta_{i}$ (with $i=1,3,4,5$) are constants. More specifically, in the absence of the linear potential $V(\phi)=m^3 \phi$, i.e., for $m=0$, a self-accelerating de Sitter solution exists that satisfies $X={\rm constant}$ \cite{DT10,DT10b,Kobayashi:2010cm,Ginf2}. Below we consider this case, for simplicity.

Thus, for covariant Galileons, taking into account $m=0$, the conditions (\ref{ttthroat}) and (\ref{rrthroat}) take the following form
\be
\frac{1}{r_0^2}= \frac{8\pi \beta_1 X_0}{1-48\pi \beta_4 X_0^2} \,, 
	\label{ttthroatGAL}
\ee
\bea
&&\frac{r''_0}{r_0} =  \frac{8\pi \beta_1  X_0}{A_0}  
	\Big[ \beta_4^2 \pi^3 (\beta_1\beta_5-\beta_3\beta_4) X_0^7 
		\nonumber  \\
	&& \qquad +\frac{\pi^3}{3} \left(\beta_1^2 \beta_5^2-4\beta_1\beta_3\beta_4\beta_5 -9\beta_1\beta_4^3 +3 \beta_3^2\beta_4^2 \right) X_0^6
	\nonumber  \\
	&&\qquad + \frac{\pi^2 \beta_4}{24}\left( \beta_1\beta_5 -\frac{5}{2}\beta_3\beta_4  \right) X_0^5 
	\nonumber  \\
	&&\qquad + \frac{3\pi^2 \beta_4}{16}\left( \beta_1\beta_4 +\frac{2}{9}\beta_3^2  \right) X_0^4 
	-\frac{\pi}{768}\left(  \beta_1\beta_5 + \beta_3\beta_4   \right) X_0^3 
	\nonumber  \\
	&&\qquad +\frac{\pi}{768}\left(  \beta_1\beta_4 - \beta_3^2   \right) X_0^2
	+\frac{\beta_3 X_0}{12288}-\frac{\beta_1}{12288} \Bigg]
	\Big/ 
		\nonumber \\ 	
	&&\qquad \Big\{\left( 48\pi\beta_4 X_0^2 -1  \right) 
	\Big[	\beta_4^2 \pi^3 (\beta_1\beta_5 +3 \beta_3\beta_4) X_0^7 
	\nonumber \\
	&&\qquad +\frac{\pi^3}{3} \left(\beta_1^2 \beta_5^2 +6\beta_1\beta_3\beta_4\beta_5 -27\beta_1\beta_4^3 
	-27 \beta_3^2\beta_4^2 \right) X_0^6
	\nonumber  \\
	&&\qquad + \frac{\pi^2 \beta_4}{24}\left( \beta_1\beta_5 +\frac{3}{2}\beta_3\beta_4  \right) X_0^5 
	\nonumber  \\
	&&\qquad - \frac{\pi^2}{24}\left( \beta_1\beta_3\beta_5 - \frac{9}{2}\beta_1\beta_4^2 -9 \beta_3^2 \beta_4  
	\right) 
		X_0^4 
	\nonumber  \\
	&& \qquad -\frac{\pi}{768}\left(  \beta_1\beta_5 +5 \beta_3\beta_4   \right) X_0^3 
	+\frac{\pi}{768}\left(  \beta_1\beta_4 - \beta_3^2   \right) X_0^2
	\nonumber \\ 
	&& \qquad +\frac{\beta_3 X_0}{12288}-\frac{\beta_1}{12288} \Bigg\}
	\,, 
\label{rrthroatGAL}
\eea
respectively. Note that not much information can be extracted from this lengthy expression, so it is useful to consider specific cases. 

For instance, consider the case of $\beta_4=0$ and $\beta_5=0$, so that Eqs. (\ref{ttthroatGAL}) and (\ref{rrthroatGAL}) reduce to 
\bea
\frac{1}{r_0^2} &=& 8\pi \beta_1 X_0 \,, 
	\label{ttthroatGALa} \\
\frac{r''_0}{r_0} &=&  \frac{8\pi \beta_1  X_0}{A_0} \left(  \frac{\beta_1- \beta_3 X_0 + 16\pi \beta_3^2 X_0^2}{\beta_1- \beta_3 X_0 + 48\pi \beta_3^2 X_0^2}  \right)\,.
	\label{rrthroatGALa}
\eea
Condition (\ref{ttthroatGALa}) imposes that $\beta_1<0$, and (\ref{rrthroatGALa}) places specific restrictions on $\beta_3$.

Second, consider the case of $\beta_3=0$ and $\beta_4=0$, so that Eqs. (\ref{ttthroatGAL}) and (\ref{rrthroatGAL}) reduce to 
\bea
\frac{1}{r_0^2} =  8\pi \beta_1 X_0 \,, \qquad \frac{r''_0}{r_0} =  \frac{8\pi \beta_1  X_0}{A_0}\,,
	\label{ttthroatGALb} 
\eea
which impose that $\beta_1<0$, in order to have wormhole geometries.

Third, consider the case of $\beta_3=0$ and $\beta_5=0$, so that Eqs. (\ref{ttthroatGAL}) and (\ref{rrthroatGAL}) reduce to 
\bea
\frac{1}{r_0^2} &=& \frac{8\pi \beta_1 X_0}{1-48\pi\beta_4 X_0^2} \,, 
	\label{ttthroatGALc} \\
\frac{r''_0}{r_0} &=&  \frac{8\pi \beta_1  X_0}{A_0} \frac{1-16\pi\beta_4 X_0^2}{\left(1-48\pi\beta_4 X_0^2\right)^2}\,.
	\label{rrthroatGALc}
\eea
Now, if $\beta_4<0$, both conditions above impose that $\beta_1<0$, to have wormhole geometries. If $\beta_4>0$, then we have two retrictions, namely: (i) the conditions $\beta_1<0$ and $\beta_4>1/(48\pi X_0^2)$ are imposed, or (ii) $\beta_1>0$  and $\beta_4<1/(48\pi X_0^2)$, to have wormhole solutions. Furthermore, if $\beta_4=0$, this reduces to $\beta_1<0$, and to the phantom case without a potential, where we identify $\beta_1=\epsilon$.

We note that wormhole geometries have been considered for a specific subclass of a Galileon Lagrangian given by ${\cal L}=F(\phi,X)+K(\phi,X) \Box\phi$, i.e., $G_2=F(\phi,X)$ and $G_3=-K(\phi,X)$ \cite{Rubakov:2015gza,Rubakov:2016zah,Kolevatov:2016ppi}. For this specific subclass, it was argued that these theories do not admit stable, static and spherically symmetric asymptotically flat traversable wormholes. Our analysis further generalizes the Lagrangian considered in \cite{Rubakov:2015gza,Rubakov:2016zah,Kolevatov:2016ppi}. Indeed, we have found the specific conditions, at the throat, that these more general subclasses of theories will allow the existence of wormhole geometries, and will hopefully spur research in this context.

\subsection{Gauss-Bonnet couplings}
 
An interesting modified gravitational theory is the Gauss-Bonnet coupling given by the action \cite{NOS,Mota,Mota2,TS06}
\be
{\cal S}_{\rm H}=\int d^4 x \sqrt{-g} 
\left[ \frac{1}{16\pi}R+X-V(\phi)+\xi(\phi){\cal G} \right]
\label{GBcouplingLag}
\ee
which includes a coupling of the form $\xi(\phi){\cal G}$ \cite{Lovelock}, where  $\xi(\phi)$ is a function of $\phi$ and ${\cal G}$ is the Gauss-Bonnet curvature invariant defined by 
\be
{\cal G}=R^2-4R_{\alpha \beta}R^{\alpha \beta}
+R_{\alpha \beta \gamma \delta}R^{\alpha \beta \gamma \delta}\,.
\ee

These theories can be accommodated in the framework of Horndeski theories 
for the following choice of the factors \cite{KYY}
\bea
G_2&=& \epsilon X-V(\phi)+8\xi^{(4)} (\phi) X^2 \left( 3-\ln (-X)  \right)\,,
	\non \\
G_3&=&-4\xi^{(3)} (\phi) X \left( 7-3\ln (-X) \right)\,,
	\non \\
G_4&=&\frac{1}{16\pi}+4\xi^{(2)}(\phi)X \left( 2-\ln (-X) \right)\,,
	\non\\
G_5&=&-4\xi^{(1)} (\phi) \ln X\,,
\eea
where $\xi^{(n)}(\phi) \equiv \partial^n \xi (\phi)/\partial \phi^n$.

Now, Eq. (\ref{ttthroat}) provides the following general relation:
\begin{equation}
 \frac{1}{r_0^2} = 8\pi \left[ \epsilon X_0 + V_0 - 48 X_0^2 \, \xi^{(4)}(\phi_0) \left( \ln|X_0| -\frac{7}{3}  \right)  \right] .
\end{equation}
However, Eq. (\ref{rrthroat}) yields an extremely lengthy expression, from which it is difficult to extract any useful information. Thus, we rather restrict ourselves to simple examples, such as, a zero potential $V(\phi)=0$, and two specific cases for the couplings, namely, (i) the linear coupling $\xi(\phi)=\lambda_1 \phi$, for which the theory is shift symmetric \cite{Yunes:2013dva,Sotiriou:2014pfa}, i.e., it is invariant for $\phi\rightarrow \phi +{\rm const.}$; and (ii) a quadratic function $\xi=\lambda_2 \phi^2$ \cite{Doneva:2017bvd,Silva:2017uqg}, which leads to a spontaneous scalarization of black holes, i.e., to dynamical formation of nonperturbative scalar field configurations.

For the linear case, conditions (\ref{ttthroat}) and (\ref{rrthroat}) simplify to
\bea
\frac{1}{r_0^2} = 8\pi \epsilon X_0,  \qquad  \frac{r''_0}{r_0} = \frac{8\pi \epsilon X_0}{A_0} \,, 
	\label{ttthroatGBcoupa} 
\eea 
where both conditions are satisfied for the case $\epsilon=-1$.

For the quadratic case $\xi=\lambda_2 \phi^2$, Eqs. (\ref{ttthroat}) and (\ref{rrthroat}) yield
\bea
\frac{1}{r_0^2} &=& 8\pi \epsilon X_0 \,, 
	\label{ttthroatGBcoupb} \\
\frac{r''_0}{r_0} &=& \frac{8\pi \epsilon X_0}{A_0} \Bigg[ -\frac{\lambda_2^2 \pi^2 \phi_0}{8}X_0^2 \ln(-X_0) 
+  \lambda_2^2 \pi^3 \epsilon \phi_0^2 X_0^2
	\nonumber \\
&&  + \frac{\lambda_2^2 \pi^2 X_0^2 (\phi_0-2)}{4}	+  \frac{\lambda_2 \pi X_0 (\phi_0-6)}{1024}	
- \frac{1}{65536} \Bigg] \Big/
	\nonumber \\
&&	\Bigg[ -\frac{\lambda_2^2 \pi^2 \phi_0}{8}X^2 \ln(-X_0) 
+ \lambda_2^2 \pi^3 \epsilon \phi_0^2 X_0^2
	\nonumber \\
&&  + \frac{\lambda_2^2 \pi^2 X_0^2 \phi_0}{4}	+  \frac{\lambda_2 \pi X_0 (\phi_0 -4)}{1024}	
- \frac{1}{65536} \Bigg] 
	  \,.
	\label{rrthroatGBcoupb}
\eea 
Note that condition (\ref{ttthroatGBcoupb}) imposes that $\epsilon=-1$, in order to have wormhole geometries. Condition (\ref{rrthroatGBcoupb}) provides one with the most general condition for this subclass of quadratic Gauss-Bonnet couplings. Wormhole geometries in the context of the Gauss-Bonnet coupling considered by the action (\ref{GBcouplingLag}) has been largely unexplored in the literature, and hopefully the analysis outlined in this subsection will spur research in wormhole geometries in this context.

\section{Conclusions}\label{S:conclusion}

A fundamental ingredient in wormhole physics is the flaring-out condition at the throat which, in classical general relativity, entails the violation of the null energy condition. However, it has been shown that in the context of modified gravity, one may impose that the matter fields threading the wormhole throat satisfy all of the energy conditions, and it is the higher order curvature terms, which may be interpreted as a gravitational fluid, that support these nonstandard wormhole geometries. Thus, it was explicitly shown that wormhole geometries can be theoretically constructed without the presence of exotic matter, but are sustained in the context of modified gravity. This has recently spurred research of wormhole physics in modified gravity. Indeed, most of these extended theories of gravity can be cast into an equivalent scalar-tensor representation. Given the large number of models, the question arises how we should study and compare them in a unified manner. 

A particularly useful tool in this direction, is the realisation that all these classes of models are special cases of the most general Lagrangian which leads to second order field equations, namely, the Horndeski Lagrangian \cite{Horndeski:1974wa}, which was recently rediscovered \cite{Deffayet:2011gz}. It enables researchers to adopt a unifying framework, and to determine subsets within this general theory that have appealing theoretical properties.
In combination with the need to fit observations such properties are helpful in preferring regions of this general theory, and hence particular models. One example of such an appealing theoretical property is the possibility that terms within the Horndeski Lagrangian can be used to partially explain the huge discrepancy between the value of the vacuum energy in particle physics, and the value of the cosmological constant as inferred from cosmological observations.

In this work, we consider the full Horndeski Lagrangian applied to wormhole geometries and present the full gravitational field equations. We analyse the general constraints imposed by the flaring-out conditions at the wormhole throat and consider a plethora of specific subclasses of the Horndeski Lagrangian, namely, quintessence/phantom fields, $k$-essence, scalar-tensor theories, covariant galileons, nonminimal kinetic coupling, kinetic gravity braiding, and the scalar-tensor representations of Gauss-Bonnet couplings and Gauss-Bonnet gravity, amongst others. 

Note that is this work, we have used the specific metric given by Eq. (\ref{metric}), however, the analysis could be generalized by considering the general line element provided by $ds^2=-f(u)dt^2+g(u)du^2+r^2(u)d\Omega^2$, where the metric functions $f(u)$, $g(u)$ and $r(u)$ are solely functions of the radial coordinate $u$. In order to avoid event horizons and singularities throughout the spacetime, one imposes that the metric functions $f(u)$ and $g(u)$ are positive and regular everywhere, and $r(u)$ also obeys the flaring-out restrictions provided by conditions (\ref{flareout}). Here, the kinetic term is given by $X=-\phi'^2/2g(u)$ and as in this work, is negative everywhere. However, the field equations are extremely lengthy and messy, and shall be considered in a future work.

The generic constraints analysed in this work serves as a consistency check of the main solutions obtained in the literature and draws out new avenues of research in considering applications of specific subclasses of the Horndeski theory to wormhole physics.

\section*{Acknowledgments}
FSNL acknowledges support from the Funda\c{c}\~{a}o para a Ci\^{e}ncia e a Tecnologia (FCT) Scientific Employment Stimulus contract with reference CEECIND/04057/2017, and the research grants No. UID/FIS/04434/2019 and No. PTDC/FIS-OUT/29048/2017. S.V.S. and R.K. are supported by the RSF grant No. 16-12-10401. Partially, this work was done in the framework of the Russian Government Program of Competitive Growth of the Kazan Federal University.


\appendix

\begin{widetext}

\section{Full gravitational field equations}\label{fieldeqs}

\subsection{Effective Einstein field equations}\label{fieldeqsEFE}

Varying the action (\ref{action}) with respect to $g_{\mu\nu}$, we obtain the following equations of motion:

$tt$--component:

\bea
\label{tt}
& \displaystyle
-A^2\phi'^4 G_{5,XX}\left(\frac{r'^2}{2r^2}A'\phi'
+\frac{r'^2}{r^2}A\phi''
\right)
+A\phi'^3 G_{5,X\phi}\left(
\frac{r'^2}{r^2}A\phi'
-\frac{2r'}{r}A\phi''
-\frac{r'}{r}A'\phi'
\right)
&
\Non
& \displaystyle
+2 G_{5,\phi\phi}\frac{r'}{r}A\phi'^3
+\phi' G_{5,X}
\left(
-\frac{A'\phi'^2}{2r^2}
-\frac{A\phi'\phi''}{r^2}
-\frac{A'^3\phi'^2}{6A}
-\frac{2}{3}A'^2\phi'\phi''
-\frac{2}{3}AA'\phi''^2
+\frac{5r'^2}{2r^2}AA'\phi'^2
\right.
&
\Non
& \displaystyle
\left.
+\frac{3r'^2}{r^2}A^2\phi'\phi''
+\frac{2r'r''}{r^2}A^2\phi'^2
\right)
+G_{5,\phi}\left(\frac{4r'}{r}A\phi'\phi''
+\frac{2r''}{r}A\phi'^2
+\frac{3r'}{r}A'\phi'^2
+\frac{r'^2}{r^2}A\phi'^2
+\frac{\phi'^2}{r^2}\right)
&
\Non
& \displaystyle
+\phi'^3 G_{4,XX}\left(
2 AA'\phi''
+A'^2\phi'
+\frac{2r'}{r}AA'\phi'
+\frac{4r'}{r}A^2\phi''
\right)
+G_{4,X\phi}\left(2A\phi'^2\phi''
+A'\phi'^3
-\frac{4r'}{r}A\phi'^3\right)
&
\Non
& \displaystyle 
-2\phi'^2\, G_{4,\phi\phi}
-2 G_{4,X}\left(
\frac{2r'}{r}A\phi'\phi''
+\frac{2r''}{r}A\phi'^2
+\frac{2r'}{r}A'\phi'^2
+\frac{r'^2}{r^2}A\phi'^2\right)
-G_{4,\phi}\left(
2\phi''
+\frac{4r'}{r}\phi'
+\frac{A'}{A}\phi'\right)
&
\Non
& \displaystyle
+G_4\left(
\frac{2}{r^2A}
-\frac{4 r''}{r}
-\frac{2 r'^2}{r^2}
-\frac{2 r'A'}{rA}
\right)
-\phi'^2\, G_{3,X}\left(\frac12 A'\phi'
+A\phi''\right)
+\phi'^2\, G_{3,\phi}
+\frac{1}{A}\, G_2=0\,,
&
\eea

$rr$-component:

\bea
\label{rr}
& \displaystyle
A'\phi'^3 G_{5,XX}\left(
\frac{1}{12}A'^2\phi'^2
+\frac13 A^2\phi''^2
+\frac13 AA'\phi'\phi''
-\frac{r'^2 }{2r^2}A^2\phi'^2
\right)
-A\phi'^4G_{5,X\phi}\left(
\frac{r'}{r}A'
+\frac{r'^2}{r^2}A
\right)
\Non
& \displaystyle 
-A'\phi' G_{5,X}\left(
\frac{A'^2}{6A}\phi'^2
+\frac{2}{3}A'\phi'\phi''
+\frac{2}{3}A\phi''^2
-\frac{5r'^2}{2r^2}A\phi'^2
+\frac{\phi'^2}{2r^2}
\right)
+\phi'^2 G_{5,\phi}\left(
\frac{3r'^2A}{r^2}
+\frac{3r'A'}{r}
-\frac{1}{r^2}\right)
\Non
& \displaystyle 
+\phi'^3G_{4,XX}\left(
A'^2\phi'
+2AA'\phi''
+\frac{2r'}{r}AA'\phi'
+\frac{2r'^2}{r^2}A^2\phi'
\right)
+\phi'^3 G_{4,X\phi}\left(
\frac{4r'}{r}A
+A'\right)
\Non
& \displaystyle 
+2\phi'^2 G_{4,X}\left(
\frac{1}{r^2}
-\frac{2r'A'}{r}
-\frac{2r'^2}{r^2}A
\right)
-\phi'G_{4,\phi}\left(
\frac{4r'}{r}
+\frac{A'}{A}\right)
+G_4\left(
\frac{2}{r^2A}
-\frac{2 r'A'}{rA}
-\frac{2 r'^2}{r^2}\right)
\Non
& \displaystyle 
-\phi'^3 G_{3,X}\left(
\frac12 A'
+\frac{2r'}{r}A
\right)
-\phi'^2 \, G_{3,\phi}
+\phi'^2 \, G_{2,X}
+\frac{1}{A} \, G_2=0\,,
&
\eea

$\theta\theta$-component:

\bea
\label{theta}
& \displaystyle
-\frac12 AA'\phi'^4G_{5,XX}\left(
\frac{r'}{r}A\phi''
+\frac{r'}{2r} A'\phi'
\right)
-\phi'^3 G_{5,X\phi}\left(
\frac14 A'^2\phi'
+\frac12 AA'\phi''
+\frac{r'}{r}A^2\phi''
\right)
&
\Non
& \displaystyle
+\phi'^3 G_{5,\phi\phi}\left(
\frac12 A'
+\frac{r'A}{r}\right)
+\phi'G_{5,X}\left(
\frac{r''}{2r}AA'\phi'^2
-\frac23 AA'\phi''^2
+\frac{r'}{r}A'^2\phi'^2
-\frac{A'^3\phi'^2}{6A}
-\frac{2}{3}A'^2\phi'\phi''\right.
&
\Non
& \displaystyle
\left.
+\frac{r'}{2r}AA''\phi'^2
+\frac{3r'}{2r}AA'\phi'\phi''
\right)
+G_{5,\phi}\left(
A'\phi'\phi''
+\frac{2r'}{r}A\phi'\phi''
+\frac{ r''}{r}A\phi'^2
+\frac12 A''\phi'^2
+\frac{A'^2\phi'^2}{2A}
+\frac{2r'}{r}A'\phi'^2\right)
&
\Non
& \displaystyle
+\phi'^3 G_{4,XX}\left(
\frac32 A'^2\phi'
+3AA'\phi''
+\frac{r'}{r} AA'\phi'
+\frac{2r'}{r}A^2\phi''\right)
&
\Non
& \displaystyle
+G_{4,X\phi}\left(
2A\phi'^2\phi''
-\frac{2r'}{r}A\phi'^3\right)
-2 \phi'^2 G_{4,\phi\phi}
-G_{4,X}\left(
A'\phi'\phi''
+\frac{2r'}{r}A\phi'\phi''
+\frac{2r''}{r}A\phi'^2\right.
&
\Non
& \displaystyle
\left.
+A''\phi'^2
+\frac{A'^2\phi'^2}{2A}
+\frac{3r'}{r}A'\phi'^2\right)
-G_{4,\phi}\left(
2\phi''
+\frac{2r'}{r}\phi'
+\frac{2 A'\phi'}{A}\right)
&
\Non
& \displaystyle
-G_4\left(
\frac{2 r''}{r}
+\frac{A''}{A}
+\frac{2 r'A'}{rA}\right)
-\phi'^2 \, G_{3,X}\left(
\frac12 A'\phi'
+A\phi''\right)
+\phi'^2 \, G_{3,\phi}
+\frac{1}{A} \, G_2 =0 \,.
&
\eea

\subsection{Scalar field equation}\label{fieldeqs_scalar}

The scalar field equation reads
\be
\nabla^\mu\left(\sum_{i=2}^5 J^i_\mu\right)=\sum_{i=2}^5 P^i,
\label{App:scalareq}
\ee
where the terms $\nabla^\mu J^i_\mu$ are given by
\bea
\nabla^\mu J^2_\mu &=& G_{2XX}A\phi'^2 \left(A\phi''+\frac12 A'\phi'\right)
-G_{2X\phi}A\phi'^2
-G_{2X} \left(A\phi''+A'\phi'+2\frac{r'}{r}A\phi'\right),
\\
\nabla^\mu J^3_\mu &=& -G_{3XX}A\phi'^3 \left(\frac12 AA'\phi'' 
+2\frac{r'}{r} A^2\phi''+ \frac14 A'^2\phi' 
+\frac{r'}{r} AA'\phi'\right)
\nonumber\\
&&
+G_{3X\phi} A\phi'^2 \left(-2A\phi''-\frac12 A'\phi' 
+2\frac{r'}{r} A\phi'\right) 
+2G_{3\phi\phi} A\phi'^2
+2G_{3\phi}\left(A'\phi'+A\phi''+2\frac{r'}{r}A\phi'\right)
\nonumber\\
&&
+G_{3X}\phi'\left(4\frac{r'}{r}A^2\phi'' +2\frac{r''}{r}A^2\phi' 
+2\frac{r'^2}{r^2}A^2\phi' +AA'\phi'' +5\frac{r'}{r}AA'\phi' 
+\frac12 AA''\phi' +\frac12A'^2\phi'\right),
\\
\nabla^\mu J^4_\mu &=& 
2G_{4X} \left( \frac{r'^2}{r^2}A^2\phi'' +\frac{r'}{r}AA'\phi''
+\frac{r''}{r}AA'\phi' +\frac{r'}{r}AA''\phi' +3\frac{r'^2}{r^2}AA'\phi' +2\frac{r'r''}{r^2}A^2\phi' +\frac{r'}{r}A'^2\phi'\right.
\nonumber\\
&&
\left. -\frac{A'\phi'}{r^2} -\frac{A\phi''}{r^2} \right)
-2G_{4X\phi}\left( 2\frac{r''}{r}A^2\phi'^2 + \frac{r'^2}{r^2}A^2\phi'^2
+4\frac{r'}{r}AA'\phi'^2 +4\frac{r'}{r}A^2\phi'\phi'' +\frac{A\phi'^2}{r^2} \right.
\nonumber\\
&&
\left. +\frac12 AA''\phi'^2 +AA'\phi'\phi'' +\frac12 A'^2\phi'^2 \right)
-A\phi'^3 G_{4X\phi\phi}\left(A'+4\frac{r'}{r} A\right)
\nonumber\\
&&
-2G_{4XX} \left(4\frac{r'}{r}A^2A'\phi'^2\phi'' 
+4\frac{r'^2}{r^2}A^3\phi'^2\phi'' +\frac{5r'}{2r}AA'^2\phi'^3 
+\frac{r'}{r}A^2A''\phi'^3 +\frac{r''}{r}A^2A'\phi'^3 \right.
\nonumber\\
&&
\left. +\frac{9r'^2}{2r^2} A^2A'\phi'^3 +2\frac{r'r''}{r^2}A^3\phi'^3 
-\frac{A^2\phi'^2\phi''}{r^2} -\frac{AA'\phi'^3}{2r^2}\right) 
\nonumber\\
&&
+2G_{4XX\phi} \left(-\frac{r'^2}{r^2}A^3\phi'^4 +2\frac{r'}{r}A^3\phi'^3\phi''
+\frac14 AA'^2\phi'^4 +\frac12 A^2A'\phi'^3\phi''\right)
\nonumber\\
&&
+G_{4XXX} \left( 2\frac{r'}{r}A^3A'\phi'^4\phi'' 
+2\frac{r'^2}{r^2}A^4\phi'^4\phi'' 
+\frac{r'}{r}A^2A'^2\phi'^5 +\frac{r'^2}{r^2}A^3A'\phi'^5\right),
\\
\nabla^\mu J^5_\mu &=&
2G_{5\phi}\left(\frac{A'\phi'}{r^2} +\frac{A\phi''}{r^2} 
-2\frac{r'r''}{r^2}A^2\phi' -\frac{r'}{r}AA''\phi' 
-\frac{r'}{r}AA'\phi'' -\frac{r''}{r}AA'\phi' -3\frac{r'^2}{r^2}AA'\phi' \right.
\nonumber\\
&&
\left. -\frac{r'}{r}A'^2\phi' -\frac{r'}{r}A^2\phi''\right)
+2A\phi'^2 G_{5\phi\phi}
\left(\frac1{r^2} -\frac{r'}{r}A' -\frac{r'^2}{r^2}A\right) 
\nonumber\\
&&
+3G_{5X}\left(\frac{r'^2}{r^2}A^2A'\phi'\phi'' -\frac{r'^2}{r^2}AA'^2\phi'^2 -\frac{r'r''}{r^2}A^2A'\phi'^2 -\frac{r'^2}{2r^2}AA''\phi'^2 
+\frac{AA'\phi'\phi''}{3r^2} \right.
\nonumber\\
&&
\left.  
+\frac{A'^2\phi'^2}{6r^2} +\frac{AA''\phi'^2}{6r^2}\right)
+G_{5X\phi}\left(5\frac{r'}{r}A^2A'\phi'^2\phi'' 
+5\frac{r'^2}{r^2}A^3\phi'^2\phi''
+2\frac{r'r''}{r^2}A^3\phi'^3 \right.
\nonumber\\
&&
\left. +\frac{r'}{r}AA'^2\phi'^3 +\frac{r'}{r}A^2A''\phi'^3
+\frac{r''}{r}A^2A'\phi'^3 +\frac{7r'^2}{2r^2}A^2A'\phi'^3
-2\frac{A^2\phi'^2\phi''}{r^2} -\frac{AA'\phi'^3}{2r^2}
\right)
\nonumber\\
&&
+\frac{r'}{r}A^2\phi'^4 G_{5X\phi\phi}\left(A' +\frac{r'}{r}A\right)
+\frac14 A\phi'^3 G_{5XX} 
\left(-\frac{A'^2\phi'}{r^2} -2\frac{AA'\phi''}{r^2} 
+9\frac{r'^2}{r^2}AA'^2\phi'
\right.
\nonumber\\
&&
\left. +2\frac{r'^2}{r^2}A^2A''\phi' +14 \frac{r'^2}{r^2}A^2A'\phi''
+4\frac{r'r''}{r^2}A^2A'\phi'  \right)
-A^2\phi'^4 G_{5XX\phi} \left( 
\frac{r'}{2r}A'^2\phi' +\frac{r'}{r}AA'\phi'' +\frac{r'^2}{r^2}A^2\phi''
\right)
\nonumber\\
&&
-\frac{r'^2}{4r^2} A^3A'\phi'^5\, G_{5XXX} (A'\phi' +2A\phi''),
\eea
respectively, and the factors $P^i$ are provided by
\bea
P^2 &=& G_{2\phi},
\\
P^3 &=& G_{3\phi\phi}\phi' -\frac12 A\phi'^2 G_{3X\phi} (2A\phi'' +A'\phi'),
\\
P^4 &=& G_{4\phi}\left(-A''-4\frac{r''}{r}A -2\frac{r'^2}{r^2}A 
-4\frac{r'}{r}A' +\frac{2}{r^2}\right)
\nonumber\\
&&
+G_{4X\phi}\left(4\frac{r'}{r}A^2\phi'\phi''
+2\frac{r'^2}{r^2}A^2\phi'^2 +AA'\phi'\phi''
+4\frac{r'}{r}AA'\phi'^2 +\frac12A'^2\phi'^2 \right),
\\
P^5 &=& G_{5\phi\phi}\left( 
\frac{r'^2}{r^2}A^2\phi'^2 +\frac{r'}{r}AA'\phi'^2 -\frac{A\phi'^2}{r^2} \right) 
+G_{5X\phi}\left(
-2\frac{r'^2}{r^2}A^3\phi'^3\phi'' -2\frac{r'}{r}A^2A'\phi'^2\phi'' \right.
\nonumber\\
&& \left.
-\frac{3r'^2}{2r^2}A^2A'\phi'^3 -\frac{r'}{r}AA'^2\phi'^3 
+\frac{A^2\phi'^2\phi''}{r^2} +\frac{AA'\phi'^3}{2r^2}
\right),
\eea
respectively.


\section{Gravitational field equations at the throat}\label{fieldeqs_throat}\label{A:fieldeqs_throat}

Here, we write out the field equations, presented in Appendix \ref{fieldeqs}, at the throat, by taking into account the conditions $r'_0=0$ and $A'_0=0$, so that Eqs. (\ref{tt}), (\ref{theta}) and the scalar field equation (\ref{App:scalareq}) reduce to the following
%
\bea
\label{tt_thr}
& \displaystyle 
\frac{r''}{r}\left(2A\phi'^2 G_{5,\phi}-4A\phi'^2 G_{4,X}-4G_4\right)
+\phi''\left(2A\phi'^2G_{4,X\phi}-\frac{A\phi'^2}{r^2} G_{5,X}-2 G_{4,\phi}
-A\phi'^2\, G_{3,X}\right)
&
\Non
& \displaystyle 
=-\frac{\phi'^2}{r^2}G_{5,\phi}
+2\phi'^2\, G_{4,\phi\phi}
-\frac{2}{r^2A}G_4
-\phi'^2\, G_{3,\phi}
-\frac{1}{A}\, G_2 \,,
&
\eea
\bea
\label{theta_throat}
& \displaystyle
\frac{ r''}{r}\left(A\phi'^2G_{5,\phi}-2A\phi'^2G_{4,X}-2G_4\right)
+A''\left(\frac12 \phi'^2 G_{5,\phi}-\phi'^2 G_{4,X}-\frac{G_4}{A}\right)
&
\Non
& \displaystyle
+\phi''\left(2A\phi'^2G_{4,X\phi}-2G_{4,\phi}-A\phi'^2 \, G_{3,X}\right)=
2 \phi'^2 G_{4,\phi\phi}-\phi'^2 \, G_{3,\phi}-\frac{1}{A} \, G_2 \,,
&
\eea
and
\bea
& \displaystyle
\frac{ r''}{r}\left(4A G_{4\phi}-4A^2\phi'^2G_{4X\phi}+2A^2\phi'^2G_{3X} \right)
+A''\left(\frac{A\phi'^2}{2r^2}G_{5X}- A\phi'^2G_{4X\phi}+\frac12 A\phi'^2 G_{3X}+G_{4\phi}\right)
&
\Non
& \displaystyle
+\phi''\left(2\frac{A}{r^2}G_{5\phi}-3\frac{A^2\phi'^2}{r^2}G_{5X\phi}
-2\frac{A}{r^2}G_{4X}+2\frac{A^2\phi'^2}{r^2}G_{4XX}-A^2\phi'^2 G_{3X\phi}
+2 A G_{3\phi}+A^2\phi'^2 G_{2XX}-A G_{2X}\right)
&
\Non 
& \displaystyle
=-3\frac{A\phi'^2}{r^2} G_{5\phi\phi}+2\frac{A\phi'^2}{r^2}G_{4X\phi}+\frac{2}{r^2}G_{4\phi}
+(\phi'-2 A\phi'^2) G_{3\phi\phi}+A\phi'^2 G_{2X\phi}+ G_{2\phi} \,,
&
\label{scalar_throat}
\eea
respectively. Note that we have removed the subscript $u_0$, denoting evaluation at the wormhole throat, from the expressions above, in order to not overload the notation.


\section{General flaring-out condition at the throat}\label{flareout_throat}


Finally, eliminating the terms $A''_0$ and $\phi''_0$, one finally arrives at the most general flaring-out condition for Horndeski wormholes, solely in terms of the scalar field $\phi$, the kinetic term $X$, the factors  $G_{i}$ and their derivatives, given by (as above, we have removed the subscript $u_0$ as not to overload the notation):
\bea
\frac{r_0''}{r_0}= \Bigg\{-2A^2 \Bigg\{ 2 \left( G_{4,X} - \frac{G_{5,\phi}}{2} \right)
\Bigg[ \frac{1}{2} \left( - G_{2,XX} + G_{3,X\phi}  \right) G_{3,\phi}
+ \left( - G_{2,X\phi} + G_{3,\phi\phi}  \right) G_{4,X\phi}
+ \left( \frac{G_{2,X\phi}}{2} - G_{3,\phi\phi} \right) G_{3,X}
	\Non 
+ G_{4,\phi\phi} \left( G_{2,XX} - G_{3,X\phi}  \right) \Big] r^4
+\Bigg[ \frac{1}{2}\left( G_{2,XX} - G_{3,X\phi}  \right) G_{5,\phi}^2
+\Bigg[\left( -G_{2,XX} + G_{3,X\phi}  \right) G_{4,X} + \left(  -\frac{G_{2,X\phi}}{2} + G_{3,\phi\phi}  \right) G_{5,X}
	\Non 
	+ \left( G_{4,XX} - \frac{3}{2}G_{5,X\phi} \right) G_{3,\phi} + 4 G_{4,X\phi}^2
	-3\left(  G_{5,\phi\phi} +G_{3,X}  \right) G_{4,X\phi} + \frac{G_{3,X}^2}{2} + \frac{3}{2}G_{5,\phi\phi}G_{3,X}
		\Non
	-2 \left( G_{4,XX} - \frac{3G_{5,X\phi}}{2} \right) G_{4,\phi\phi} \Bigg] G_{5,\phi}
	+ \Bigg[ \left( G_{2,X\phi} -2 G_{3,\phi\phi}  \right) G_{5,X}
	+ \left( -2 G_{4,XX} + 3 G_{5,X\phi}  \right) G_{3,\phi}  - 4 G_{4,X\phi}^2 
	\Non
	+ 2 \left( 3 G_{5,\phi\phi} + G_{3,X}  \right) G_{4,X\phi} - 3 G_{5,\phi\phi} G_{3,X} 
	+ 4 \left( G_{4,XX} - \frac{3}{2} G_{5,X\phi}   \right) G_{4,\phi\phi} \Bigg] G_{4,X}
	\Non
	-2 \left(  G_{4,\phi\phi} - \frac{G_{3,\phi}}{2}  \right)\left(  G_{4,X\phi} - \frac{G_{3,X}}{2}  \right) G_{5,X} 
	\Bigg] r^2
	+ \left( G_{4,XX} - \frac{3G_{5,X\phi}}{2}   \right) G_{5,\phi}^2 
	+ \Bigg[\left( - 2 G_{4,XX} + 3 G_{5,X\phi}   \right) G_{4,X}	
	\Non
	- 2 \left( G_{4,X\phi} - \frac{3G_{5,\phi\phi}}{4} - \frac{G_{3,X}}{4}   \right) G_{5,X} \Bigg] G_{5,\phi}
	+ 2 \left[  \left( G_{4,X\phi} - \frac{3G_{5,\phi\phi}}{2}   \right) G_{4,X}  
	+ \frac{1}{2}  \left( G_{4,\phi\phi} - \frac{G_{3,\phi}}{2}   \right)	G_{5,X}  \right]    G_{5,X} \Bigg\} \phi'^6
	\Non
	-2 A^2 \left(  G_{4,X} - \frac{G_{5,\phi}}{2}  \right) \left[ \left( -2 G_{4,X\phi} + G_{3,X}   \right) r^2 + G_{5,X}   
	\right] G_{3,\phi\phi} r^2 \phi'^5
	- 4A^2 \Bigg\{ \Bigg[ \Bigg( \frac{1}{2} \left(  - G_{2,XX} + G_{3,X\phi}  \right)  G_{3,\phi} 
	\Non
	+ \left(  - G_{2,X\phi} +2 G_{3,\phi\phi}  \right)  G_{4,X\phi}
	+ \left(  \frac{G_{2,X\phi}}{2} - G_{3,\phi\phi}  \right)  G_{3,X}
	+\left(  G_{2,XX} - G_{3,X\phi}  \right)  G_{4,\phi\phi}		
		  \Bigg) G_{4}
	\Non
	+ \left( G_{4,X} - \frac{G_{5,\phi}}{2}  \right) \Bigg[ - G_{3,\phi}^2
	+ \left( 2 G_{4,\phi\phi} + \frac{G_{2,X}}{2}  \right)  G_{3,\phi}  
	- G_{4,X\phi}G_{2,\phi} + \frac{1}{2} G_{2,\phi} G_{3,X}
	+ \left( G_{2,X\phi} - 2 G_{3,\phi\phi}  \right)  G_{4,\phi} 
	\Non	
	- G_{4,\phi\phi} G_{2,X} - \frac{G_{2}}{2} \left( G_{2,XX} - G_{3,X\phi}  \right) \Big] \Bigg] r^4
	+ \Bigg\{\Bigg[ \left(  \frac{G_{3,X}}{2} - G_{4,X\phi}  \right) G_{5,\phi} 
	+ \left( - G_{2,XX} + G_{3,X\phi}   	\right) G_{4,X}
	\Non
	+ \left( G_{4,\phi\phi} + \frac{G_{2,X\phi}}{2}  -  \frac{G_{3,\phi}}{2} - G_{3,\phi\phi}	\right) G_{5,X}
	+ \left(  G_{4,XX} - \frac{3G_{5,X\phi}}{2} 	\right) \left( 2 G_{4,\phi\phi}  - G_{3,\phi} \right)
	+ \left( - G_{3,X} + 3G_{5,\phi\phi} 	\right) G_{4,X\phi} 
		\Non	
	+ \frac{1}{2} G_{3,X}^2 - \frac{3}{2}G_{5,\phi\phi}G_{3,X} \Bigg] G_4
	+ \left( - \frac{G_{2,X\phi}}{4} + G_{3,\phi} - G_{4,\phi\phi}  \right)  G_{5,\phi}^2 
	+ \Bigg[  \left( 3 G_{4,\phi\phi} + \frac{G_{2,X}}{2} - \frac{5G_{3,\phi}}{2}  \right)  G_{4,X} 
	- \frac{G_{2,\phi} G_{5,X} }{4}
	\Non
	- G_{4,X\phi} G_{4,\phi}  + \frac{3}{2} G_{4,\phi} G_{5,\phi\phi}  
	+  \frac{1}{2} \left(  G_{4,XX} - \frac{3}{2} G_{5,X\phi}    \right) G_2 \Bigg]  G_{5,\phi} 
	+ \left(  G_{3,\phi} - 2 G_{4,\phi\phi}    \right)  G_{4,X}^2 
	+ \Bigg[ \frac{1}{2} G_{2,\phi} G_{5,X}  +  G_{3,X}  G_{4,\phi}
		\Non	
	 -  3 G_{4,\phi}  G_{5,\phi\phi} 
	- \left(  G_{4,XX} - \frac{3}{2} G_{5,X\phi}    \right) G_2  \Bigg] G_{4,X}
	+ \frac{G_2}{2} \left(  G_{4,X\phi} - \frac{1}{2} G_{3,X}    \right) G_{5,X} \Bigg\}r^2
	+ \Bigg[ \left( -2 G_{4,XX} + 3 G_{5,X\phi}    \right) G_{4,X}
	\Non
	- \frac{3}{2} \left( - \frac{G_{3,X}}{3} +  G_{5,\phi\phi}    \right) G_{5,X}  \Bigg] G_{4}
	+ \frac{G_{5,\phi}^3}{2} +  \left( - \frac{3  G_{5,\phi}}{2} +  G_{4,X}   \right)  G_{4,X} G_{5,\phi}
	+G_{5,X} \left( -\frac{G_{2}G_{5,X}}{4} +  G_{4,X} G_{4,\phi}    \right)  \Bigg\} \phi'^4
	\Non
	-4 A G_{3,\phi\phi} r^2 \left\{ \left[  \left( \frac{G_{3,X}}{2} -   G_{4,X\phi}    \right)  G_4
	+ \left(  G_{4,X}  -\frac{G_{5,\phi}}{2}     \right)   G_{4,\phi} 	 \right] r^2 
	+ \frac{G_{5,X} }{2}  G_{4,X}  \right\} \phi'^3
		\Non	
	- 4 A  \Bigg\{ \Bigg[ \Big[ -  G_{3,\phi}^2 
	 +\left( 2 G_{4,\phi\phi}  + \frac{G_{2,X}}{2}     \right)   G_{3,\phi}
	- G_{4,X\phi}  G_{2,\phi} + \frac{ G_{2,\phi}  G_{3,X}}{2}
	+ \left( G_{2,X\phi}  - 2 G_{3,\phi\phi}     \right)   G_{4,\phi}
	- G_{4,\phi\phi}  G_{2,X}   
	\Non
	- \frac{1}{2}  \left( G_{2,XX}  - 3 G_{3,X\phi}     \right)   G_{2} \Big]     G_{4}
	+ \left( \frac{1}{2}  G_{2}  G_{2,X} - G_{2}  G_{3,\phi} + G_{4,\phi}  G_{2,\phi}   \right) 
	\left(  G_{4,X}  -  \frac{G_{5,\phi}}{2} \right) \Bigg] r^4  
	\Non
	+ \Bigg[  \left( -  G_{2,XX} - 2 G_{4,X\phi} + G_{3,X\phi} + G_{3,X}   \right)  G_{4}^2 
	+ \Big[  \left( -  G_{3,\phi} + 2 G_{4,\phi\phi} + G_{4,\phi}   \right)  G_{5,\phi}
	+  \left( -  G_{3,\phi} - 2 G_{4,\phi\phi} + G_{2,X}   \right)  G_{4,X} 
	\Non
	+ \frac{1}{2} \left( - G_{2} + G_{2,\phi}   \right)  G_{5,X}  - 2 G_{4,X\phi} G_{4,\phi} 
	+ 2 G_{3,X}G_{4,\phi} - 3 G_{4,\phi} G_{5,\phi\phi}
	- \left(  G_{4,XX}  -  \frac{3 G_{5,X\phi}}{2} \right)  G_2  \Bigg]  G_4
	\Non
	+ \left(  G_{2} G_{4,X}  -  G_{2} G_{5,\phi}  + 2 G_{4,\phi}^2   \right) 
	\left( G_{4,X} - \frac{G_{5,\phi}}{2}  \right)  \Bigg]  r^2
	+ 2 \Bigg[  \left( - G_{4,XX} - \frac{3G_{5,X\phi}}{2}  \right) G_4 + G_{4,X}^2
	-  G_{4,X} G_{5,\phi} 
	\Non	
	+ G_{4,\phi}  G_{5,X} \Bigg] G_4  \Bigg\}  \phi'^2
	-  4   G_{3,\phi\phi}  G_{4} G_{4,\phi}  \phi'  r^4
	- 4 \Bigg[ \left( \frac{1}{2} G_{2} G_{2,X}  -  G_{2} G_{3,\phi}  +  G_{4,\phi}G_{2,\phi} \right) r^4
	\Non
	\left[ \left( 2 G_{4,\phi}  + G_{2,X}  -  2 G_{3,\phi} \right)  G_4    
	+   G_{2}  \left( G_{4,X}  -  G_{5,\phi}    \right)  +  2 G_{4,\phi}^2  \right]    r^2
	+   2 G_{4}  \left( G_{4,X}  -  G_{5,\phi}    \right)  \Bigg]  G_4 \Bigg\}  \Bigg/
	\Non
	\Bigg\{  8A \left[  A \left(  G_{4,X}  - \frac{1}{2} G_{5,\phi}  \right) \phi'^2  +  G_4   \right]
	\Bigg\{ A^2 \Bigg[ \left[    \left( G_{2,XX}  -  G_{3,X\phi}  \right)  
	\left( G_{4,X}  - \frac{1}{2} G_{5,\phi}  	\right)
	- 3  \left( G_{4,X\phi}  - \frac{1}{2} G_{3,X}  	\right)^2  \right] r^4
	\Non
	\left[    \left( G_{4,XX}  -  \frac{3}{2} G_{5,X\phi}  \right)  
		\left( 2 G_{4,X}  -  G_{5,\phi}  \right)
	+   \left( G_{4,X\phi}  - \frac{1}{2} G_{3,X}  	\right) G_{5,X}  \right] r^2
	+ \frac{G_{5,X}^2}{4}  \Bigg]  \phi'^4
	\Non
	- 2  \Bigg\{ \Bigg[  \frac{1}{2} \left(  - G_{2,XX}  +  G_{3,X\phi}   - G_{4,X\phi}   
	+ \frac{1}{2} G_{3,X}   \right) 	G_4
	+  \left( \frac{1}{2} G_{2,X} - G_{3,\phi}  \right)  \left( G_{4,X}  - \frac{1}{2} G_{5,\phi}  \right)	
		\Non	
	-  \frac{5}{2} G_{4,\phi} \left(  G_{4,X\phi} - \frac{G_{3,X}}{2} \right)   \Bigg] r^2
	+ \frac{1}{2}  \left(  -  \frac{G_{5,X}}{2} - 2 G_{4,XX} + 3 G_{5,X\phi} \right)  G_4
	+  G_{4,X}^2  -  \frac{3 G_{4,X} G_{5,\phi}}{2}  +   \frac{3 G_{4,\phi} G_{5,X}}{4}
	\Non
	+  \frac{ G_{5,\phi}^2}{2}  \Bigg\} A r^2 \phi'^2
	-2 \left\{  \left[ \left(  \frac{G_{4,\phi}}{2} + \frac{G_{2,X}}{2} - G_{3,\phi} \right)  G_4
	+  G_{4,\phi}^2 \right] r^2  +   \left( G_{4,X} - G_{5,\phi}  \right) G_4 \right\} r^2 \Bigg\}  \Bigg\}
	\Non
	\label{rrthroat}
\eea

Note that this relation is constrained by the imposition of the flaring-out condition, $r''_0>0$, at the throat. Thus, in order to be a wormhole solution, this equation, in addition to $r_0>0$ given by condition (\ref{ttthroat}), impose tight restrictions on the spacetime geometry.

\end{widetext}



\end{document}